\documentclass[aps,nofootinbib,11pt,preprintnumbers]{revtex4}
\usepackage{amssymb,amsmath}
\usepackage{hyperref}
\newcommand{\bea}{\begin{eqnarray}}
\newcommand{\eea}{\end{eqnarray}}

\def\x{{\bf x}}
\def\k{{\bf k}}
\def\H{{\cal H}}
\def\tr{{\rm tr}}
\def\volume{{\cal V}}
\def\eF{\varepsilon_{\scriptscriptstyle F}}

\begin{document}
\preprint{MIT-CTP 3978}

\title{
Black holes and non-relativistic quantum systems
}
\author{Pavel Kovtun}
\affiliation{Department of Physics and Astronomy,
University of Victoria, Victoria BC, V8P 5C2, Canada}
\author{Dominik Nickel}
\affiliation{Center for Theoretical Physics,
  Massachusetts Institute of Technology,
  Cambridge, MA 02139, USA}
\date{September 2008}

\begin{abstract}
\noindent
We describe black holes in $d+3$ dimensions, 
whose thermodynamic properties correspond to those of
a scale invariant non-relativistic $d+1$ dimensional quantum system
with dynamical exponent $z=2$.
The gravitational model involves a massive abelian vector field
and a scalar field, in addition to the metric.
The energy per particle in the dual theory is $|\mu|\, d/(d{+}2)$
at any temperature,
while the ratio of shear viscosity to entropy density is
$\hbar/4\pi$. 
\end{abstract}

\maketitle

\section{Introduction}
\noindent
Scale invariance is an important concept in physics,
with fundamental applications in diverse areas including
critical phenomena and high energy physics.
One can think about scale invariance as a symmetry under
simultaneous rescaling of both space and time
$x^i\to\lambda x^i$, $t\to\lambda^z t$, where $\lambda>0$ is the
scaling parameter, and $z$ is the dynamical exponent.
Relativistic scale invariant systems such as blackbody radiation
have $z=1$,
while many non-relativistic scale-invariant systems have $z=2$.
An interacting non-relativistic quantum system which has recently attracted
much interest is given by a collection of cold fermionic atoms in the
unitary limit \cite{cold-atoms-review}.
In this limit the scattering length of atoms formally diverges whereas the
mean particle distance is considered (arbitrary) large compared to the range of
the interaction.
The only relevant length scales are then given by the mean particle distance
and the thermal wavelength, which are determined through the fermion density or
chemical potential and the temperature.
All quantities can therefore be expressed by their analogues in the
non-interacting theory times a universal function only depending on the
dimensionless ratio of chemical potential and temperature~\cite{Ho:2004}.
Consequently, the system at zero temperature and density should be
described in terms of a non-relativistic scale invariant theory
with $z=2$~\cite{Mehen:1999nd}.

The methods of the gauge-gravity duality \cite{Aharony:1999ti}
have been extremely helpful for
understanding strongly interacting scale invariant relativistic models, and
it is reasonable to ask whether the same tools can be useful in understanding
strongly interacting scale invariant non-relativistic models.
The key idea lies in realizing the relevant symmetry group geometrically,
rather than directly on the Hilbert space of the quantum theory.
The algebra of the non-relativistic conformal group (the Schr\"odinger group)
can be obtained from the relativistic group by reducing along the light-cone and
identifying the corresponding momentum with the particle number
operator~\cite{Son:2008ye}. A spacetime realizing this symmetry as its
isometry group was recently identified in~\cite{Son:2008ye}
for $z=2$, and independently in \cite{Balasubramanian:2008dm} for arbitrary $z$:
\bea
\label{eq:ds20}
ds^2 &=& - r^{2z} du^2 - 2r^2 dudv + r^2 dx^i dx^i +
\frac{dr^2}{r^2}
\,.
\eea
Here $r$ is the radial coordinate in the standard gauge-gravity duality
description with the
boundary at $r{\to}\infty$, $u$ is the boundary time coordinate,
$x^i$ are $d$-dimensional spatial coordinates, and $v$ is an additional
coordinate whose conserved momentum corresponds to the particle number operator.
For $d=2$ and $z=2$, this spacetime and its non-extremal generalizations were
embedded in type IIB string
theory~\cite{Adams:2008wt,Herzog:2008wg, Maldacena:2008wh} and the
equation of state in the dual field theory together with 
its shear viscosity were determined.
In addition, the authors of \cite{Herzog:2008wg, Maldacena:2008wh}
found a five-dimensional
gravitational model which admits a non-extremal version of (\ref{eq:ds20})
as a solution, corresponding to a $2{+}1$ dimensional scale-invariant field
theory at non-zero temperature and density.%
\footnote{%
     Other recent work on holographic realization of non-relativistic theories includes
     \cite{Goldberger:2008vg,Barbon:2008bg,Kachru:2008yh}.
}

Given that most experiments on cold atomic gases are performed in
$d=3$ spatial dimensions,
it would be interesting to find a simple gravitational model, corresponding
to a $3{+}1$ dimensional scale invariant quantum system
at non-zero temperature and density.
Such six-dimensional gravitational action would be a natural starting point
for phenomenological holographic models of non-relativistic systems, 
similar to the AdS/QCD program \cite{Erlich:2005qh}.
It is the purpose of this note to present black hole spacetimes
whose thermodynamic properties
correspond to those of a scale-invariant non-relativistic quantum system
in arbitrary spatial dimension $d\geqslant2$, 
including the physically interesting case $d=3$.
Using the standard
dictionary of the gauge-gravity duality, we calculate the thermodynamic
potentials and the shear viscosity of the dual $d{+}1$ dimensional
non-relativistic field theory.
Our notations closely follow the presentation of Herzog et al~\cite{Herzog:2008wg}.

%%%%%%%%%%%%%%%%%%%%%%%%%%%%%%%%%%%%%%%%%%%%%%%%%%%%%%%%%%%%%%%%%%%%%%
\section{Thermodynamics of a scale-invariant system}
\noindent
Before presenting black hole spacetimes whose thermodynamics
corresponds to that of a scale invariant system, let us briefly
discuss what we expect thermodynamic potentials to look like.
We are interested in a translation invariant
quantum field theory in $d{+}1$ spacetime dimensions
in the thermodynamic limit in the grand canonical
ensemble given by the density operator $\rho=Z^{-1}\exp[-(H{-}\mu N)/T]$
where $Z=\tr\exp[-(H{-}\mu N)/T]$ is the partition function,
$T$ is temperature, $H$ is the Hamiltonian,
$\mu$ is the chemical potential, and
$N$ is the conserved particle number operator which commutes with $H$.
Scale invariance means that there is a time-independent operator $D$
whose commutation relations within the Schr\"odinger algebra include%
\footnote{%
     The second relation in (\ref{eq:D-comm}) holds in systems where
     particle number current coincides with momentum density.
     From a geometric perspective, this commutation relation
     emerges from the isometries of the spacetime (\ref{eq:ds20})
     \cite{Balasubramanian:2008dm}.
     We thank J.~McGreevy for a discussion on these points.
}
\begin{equation}
\label{eq:D-comm}
   [D,H] = izH\,, \ \ \
   [D,N] = i(2{-}z)N\,.
\end{equation}
The Hamiltonian density $\H$ has scaling dimension $d{+}z$, i.e.
$[D,\H(0)] = i(d{+}z)\H(0)$.
Taking the expectation value of this equation in the grand canonical
ensemble, using the commutation relations (\ref{eq:D-comm}) and cyclicity of the trace gives
$$
   \left(
   z\,T \frac{\partial}{\partial T} - 
   2(1{-}z)\,\mu\frac{\partial}{\partial\mu}
   \right)
   \epsilon = (d{+}z)\epsilon\,,
$$
where $\epsilon=\tr(\rho\H)$ is the equilibrium energy density.
The solution  is 
$\epsilon = T^{\frac{d+z}{z}} g(T^{\frac{2-2z}{z}}\mu)$ 
where $g$ is an arbitrary function.
Once the energy density is known,
pressure $p$ can be determined from the relation $\epsilon+p=Ts+\mu n$,
where $s=\partial p/\partial T$ is entropy density,
and $n=\partial p/\partial\mu$ is the number density.
Demanding that $p{\to}0$ as $\epsilon{\to}0$, one finds for $z=2$
\begin{equation}
\label{eq:thermodynamics}
   \epsilon(T,\mu)=T^{\frac{d+2}{2}}g\!\left({\mu}/{T}\right),\ \ \ 
   2\,\epsilon(T,\mu) = d\, p(T,\mu)\,.
\end{equation}
We will now describe black holes in $d{+}3$ dimensions
whose thermodynamics has exactly the
scaling form of Eq.~(\ref{eq:thermodynamics}),
with a particular function $g(\mu/T)$.

%%%%%%%%%%%%%%%%%%%%%%%%%%%%%%%%%%%%%%%%%%%%%%%%%%%%%%%%%%%%%%%%%%%%%%
\section{Gravitational model}
\label{sec:model}
\noindent
We propose the following action of gravity coupled to a
massive abelian vector field plus a scalar:
\bea
\label{eq:S1}
   \mathcal{S} &=& \frac{1}{16\pi G_{d+3}}
   \int\!\! d^{d+3}x\, \sqrt{-g}
   \left( R -
   \frac{a}{2}(\partial_\mu\phi)(\partial^\mu\phi) -
   \frac{1}{4} e^{-a\phi}F_{\mu\nu}F^{\mu\nu} -
   \frac{m^2}{2}A_\mu A^\mu - V(\phi)
   \right) \,,
\eea
where $G_{d+3}$ is the $(d{+}3)$-dimensional Newton's constant,
the scalar potential is given by 
$ V(\phi) = (\Lambda{+}\Lambda')e^{a\phi} + (\Lambda{-}\Lambda')e^{b\phi}$,
and the coefficients are 
$$
   \Lambda=-\frac{1}{2}(d{+}1)(d{+}2) \,,\quad
   \Lambda'=\frac{1}{2}(d{+}2)(d{+}3) \,,\quad
   m^2=2(d{+}2) \,,\quad 
   a=(d{+}2)b=2\frac{d{+}2}{d{+}1} \,.
$$
For $\phi=0$, one recovers the model of refs.~\cite{Son:2008ye,Balasubramanian:2008dm},
where the spacetime (\ref{eq:ds20}) emerges as a solution to Einstein equations
with a non-trivial profile of $A_\mu$.
For $\phi\neq0$, our model admits the following planar black hole solutions
which are asymptotic to (\ref{eq:ds20})
\begin{subequations}
\label{eq:bh-uv}
\bea
    && ds^2 = r^2 h^{-\frac{d}{d+1}} \left(
    \left[ \frac{(f{-}1)^2}{4(h{-}1)} - f \right] r^2 du^2 - 
    (1{+}f) dudv + \frac{h{-}1}{r^2}dv^2 \right) + 
    h^{\frac{1}{d+1}}
    \left( r^2 dx^i dx^i + \frac{dr^2}{r^2 f} \right),\ \ \ \ \\
    && A = \frac{1{+}f}{2h}r^2 du + \frac{1{-}h}{h}dv\,,\\
    && \phi = -\frac12 \ln h\,,
\eea
\end{subequations}
where
$ h(r) = 1+{\beta^2r_0^{d+2}}/{r^{d}}$ and $ f(r) = 1-{r_0^{d+2}}/{r^{d+2}}$.
Here $\beta$ is an arbitrary parameter, 
and the horizon is at $r=r_0$.
It is helpful to introduce coordinates
$t= \beta v + {u}/(2\beta)$ and
$y=-\beta v + {u}/(2\beta)$, so that the solution (\ref{eq:bh-uv}) becomes
\begin{subequations}
\label{eq:bh-ty}
\begin{eqnarray}
    && ds^2 = r^2 h^{-\frac{d}{d+1}} 
    \left(-f dt^2 + dy^2 -r^2 f\beta^2 (dt{+}dy)^2 + h\, dx^i dx^i \right) + 
    h^{\frac{1}{d+1}}  \frac{dr^2}{r^2 f} \,,\ \ \ \ \\
    && A(r) = \frac{r^2\beta}{h}(f dt + dy)\,.
\end{eqnarray}
\end{subequations}
For $d{=}2$ these solutions were discussed in ref.~\cite{Herzog:2008wg}.
Note that Eq.~(\ref{eq:bh-uv}) with $f(r)=1$, $h(r)=1+\Omega^d/r^d$
is also a solution, with $\Omega$ a free parameter.
We will now show that thermodynamics of black holes (\ref{eq:bh-uv}), (\ref{eq:bh-ty})
matches thermodynamics of $d{+}1$ dimensional scale-invariant field theories with $z=2$.

%%%%%%%%%%%%%%%%%%%%%%%%%%%%%%%%%%%%%%%%%%%%%%%%%%%%%%%%%%%%%%%%%%%%%%%%%%%%
\section{Black hole thermodynamics}
\noindent
The Bekenstein-Hawking entropy of the black hole is proportional to the
area of the horizon,
\begin{equation}
\label{eq:bh-entropy}
     S = \frac{1}{4G_{d+3}}r_0^{d+1}\beta\,\Delta v\,\volume\,,
\end{equation}
where we assume that $v$ has a finite size $\Delta v$, and
$\volume=\Delta x^1\dots\Delta x^d$ is the (infinite) volume
along the $x^i$ directions which we identify with the 
spatial volume of the dual field theory. 
The temperature of the black hole can be found 
from the surface gravity $\kappa$,
\begin{equation}
  \kappa^2 = -\frac{1}{2} (\nabla^\mu \xi^\nu)(\nabla_{\!\mu}\xi_\nu) \,,
\end{equation}
where $\xi$ is a Killing vector field which is null at the horizon.
Following \cite{Herzog:2008wg}, we choose $\xi^t=1/\beta$, or
\bea
\label{eq:xi}
    \xi=\frac{\partial}{\partial u} + \frac{1}{2\beta^2} \frac{\partial}{\partial v}\,.
\eea
This gives the black hole temperature $T=\kappa/2\pi$ equal to
\begin{equation}
\label{eq:bh-temperature}
   T = \frac{(d{+}2)r_0}{4\pi\beta}\,,
\end{equation}
which we identify as the temperature of the dual field theory.
 From (\ref{eq:xi}) we can also infer the value of the chemical potential. With
the generator of translations in the $u$-direction corresponding to the
Hamiltonian and the generator of translations in the $v$-direction
corresponding to particle number, we expect the chemical
potential to be \cite{Herzog:2008wg} 
\begin{equation}
     \mu = -\frac{1}{2\beta^2} 
\end{equation}
as the conjugate Lagrange multiplier to the particle number
(this is analogous to what happens in thermodynamics of Kerr black holes).
With this identification, the chemical potential is negative.

We identify the partition function as $Z=e^{-{\cal S}_E}$
where ${\cal S}_E$ is the the on-shell Euclidean action,
obtained after taking $t\to i\tau$.
The on-shell action is divergent at large $r$, and requires regularization.
To make the action finite, we add counterterms at the boundary $r{\to}\infty$,
\begin{equation}
\label{eq:Sreg}
     \mathcal{S} \rightarrow \mathcal{S} + \frac{1}{8\pi G_{d+3}} \int\!\!d^{d+2}x\,
     \sqrt{-h} \left(K - c_0 + \dots\right)\,,
\end{equation}
where $K$ is the standard Gibbons-Hawking boundary term,
$c_0$ is the boundary cosmological constant,
and dots denote ``matter'' counterterms which are polynomial in the fields
$\phi$, $A_\mu$ and their derivatives.
The on-shell action evaluated on the solution (\ref{eq:bh-uv}) 
can be made finite by choosing $c_0=d{+}1$,
together with the appropriate coefficients 
of the matter counterterms, giving%
\footnote{%
     Alternatively, the Euclidean on-shell action can be regularized by
     subtracting the action of the zero-temperature solution (\ref{eq:bh-uv})
     with $f(r)=1$ and $h(r)=1+\beta^2 r_0^{d+2}/r^d$, and demanding that
     the volumes at large fixed $r$ coincide in the two geometries.
     This procedure gives the same answer (\ref{eq:action-regularized}).
}
\begin{equation}
\label{eq:action-regularized}
     \mathcal{S}_E = -\frac{1}{16\pi G_{d+3}} \int\!\! d^{d+2}x \; r_0^{d+2} \,.
\end{equation}
The matter counterterms do not contribute to the
finite part of the action for $d>2$.
In order to evaluate the thermodynamic potentials,
the Euclidean ``time'' direction $\tau$ 
is compactified on a circle with period 
$\Delta\tau={4\pi}/{(d{+}2)r_0}={1}/(\beta T)$.
The grand canonical potential $\Omega=-p\volume=-T\ln Z$ is then given by
\begin{equation}
\label{eq:Omega}
    \Omega(T,\mu) = -\tilde{c}\,\volume\, T^{\frac{d+2}{2}}
                    \left(\frac{T}{|\mu|}\right)^{\!\frac{d+2}{2}}\,,
\end{equation}
where the $(T,\mu)$-independent constant
$\tilde{c} = 
 (\Delta v) \pi^{d+1} 2^{(3d-2)/2}(d{+}2)^{-(d+2)}/G_{d+3}$
determines the normalization of all thermodynamic functions, and therefore
effectively counts the degrees of freedom in the dual field theory.%
\footnote{%
    In the analogous gravitational description of relativistic field theories,
    the normalization coefficient $\tilde{c}$ is uniquely related
    to the central charge of the corresponding CFT \cite{Kovtun:2008kw}.
}
All other thermodynamic functions can be easily determined
as $n=\partial p/\partial\mu$, $s=\partial p/\partial T$,
$\epsilon = -p+Ts+\mu n$, and one finds $2\epsilon = d\,p$, 
as is expected in a scale-invariant theory with $z=2$.
The dependence on temperature and chemical potential is precisely
of the expected form (\ref{eq:thermodynamics}), with the scaling function
$g(x) = \frac{d\,\tilde{c}}{2}\, |x|^{-(d+2)/2}$.
Note that negative $\mu$ is required so that the density 
$n=\partial p/\partial\mu$ is positive.
The total entropy
\begin{equation}
     S = -\frac{\partial\Omega}{\partial T} = -\frac{d{+}2}{T}\Omega =
     -(d{+}2) {\cal S}_E  = \frac{1}{4 G_{d+3}} r_0^{d+1} \beta\Delta v\,\volume
\end{equation}
is equal to the Bekenstein-Hawking entropy (\ref{eq:bh-entropy}),
a consistency check for thermodynamics.

%%%%%%%%%%%%%%%%%%%%%%%%%%%%%%%%%%%%%%%%%%%%%%%%%%%%%%%%%%%%%%%%%%%%%%%%%%%%
\section{Viscosity}
\noindent
In a translation-invariant theory with a hydrodynamic regime at long distances,
shear viscosity can be evaluated using the Kubo formula
\begin{equation}
\label{eq:Kubo-formula}
     \eta = -\lim_{\omega\rightarrow 0}\,\frac{1}{\omega}\,
     {\rm Im}\, G^{R}_{12,12}(\omega,{\bf k}{=}0)\,,
\end{equation}
where $G^{R}_{12,12}$ is the retarded function of the component $T_{x^1 x^2}$
of the energy-momentum tensor.
The relevant component of the metric perturbation in the bulk $\varphi\equiv h_1^2$
can be Fourier transformed along the $u$ and $x^i$ directions,
$\varphi(u,v,\x,r)=\int\!d\omega\, d^dk/(2\pi)^{d+1}
 \varphi_{\omega,\k}(r)e^{-i\omega u + i\k\cdot\x}$ 
(note that $u$ is identified with the time in the dual field theory),
and only $\k{=}0$ is needed to find the viscosity.
The perturbation is $v$-independent, as required by the fact that the
energy-momentum tensor has zero particle number.
The field $\varphi$ decouples from other perturbations, 
and satisfies the equation of the minimal massless scalar \cite{Kovtun:2004de}.
In the black hole background (\ref{eq:bh-uv}), one finds
\begin{equation}
    \varphi_\omega''(r) + 
    \frac{(d{+}3)r^{d+2}-r_0^{d+2}}{r(r^{d+2}-r_0^{d+2})} \varphi_\omega'(r) +
    \frac{\omega^2\beta^2\, r^{d-2}\,r_0^{d+2}}{(r^{d+2}-r_0^{d+2})^2} 
    \varphi_\omega(r) =0\,.
\end{equation}
The equation has a regular singular point at the horizon, with exponents
$\alpha_\pm=\pm \frac{i\omega}{4\pi T}$.
Evaluation of the retarded function requires that one keeps only
the wave going into the black hole \cite{Son:2002sd,Herzog:2002pc},
which fixes the solution as
$\varphi_\omega(r)=(r{-}r_0)^{\alpha_-}F(r)$, where $F(r)$ is regular at the horizon $r=r_0$.
In the hydrodynamic limit $\omega\to0$, the equation for $F(r)$ can be solved
as a power expansion in $\omega$. To linear order, one finds
\begin{equation}
\label{eq:F-solution}
     F(r) = F_0-\frac{i\omega\beta F_0}{(d{+}2)\,r_0}\ln\left(
     \frac{r_0\, f(r)}{r-r_0}\right)
     +O(\omega^2)\,,
\end{equation}
where $F_0$ is an integration constant which determines the overall normalization
of the perturbation. 
It is proportional to the boundary value of the field
$\varphi_0(\omega)\equiv\varphi_\omega(r_\Lambda)$, where $r_\Lambda{\to}\infty$
is the large-$r$ cutoff.
The two-point function is determined from the regularized on-shell action
(\ref{eq:Sreg}).
The Gibbons-Hawking term ensures that the action for $\varphi$ has the form
(up to contact boundary terms)
\begin{eqnarray}
(16\pi G_{d+3})\,
{\cal S} &=& -\frac12 \int\!\! d^{d+3}x\,
             \sqrt{-g}\, g^{\mu\nu}\partial_\mu\varphi\, \partial_\nu\varphi \nonumber\\
         &=& - \frac{1}{2} \int\frac{d\omega}{2\pi} \frac{d^d k}{(2\pi)^d}\, 
             r^{d+3}f(r)\, \varphi_{-\omega}(r)\,
             \varphi'_{\omega}(r)\, \Delta v \big|_{r=r_\Lambda}\,.
\end{eqnarray}
Taking the second variation of the action with respect to $\varphi_0(\omega)$
\cite{Son:2002sd,Herzog:2002pc},
one finds the viscosity from the Kubo formula (\ref{eq:Kubo-formula}),
using the solution (\ref{eq:F-solution}),
\begin{equation}
     \eta = \frac{r_0^{d+1}\beta \Delta v}{16\pi G_{d+3}}\,.
\end{equation}
Comparing with the expression for the black hole entropy (\ref{eq:bh-entropy}),
the ratio of shear viscosity to entropy density $s=S/\volume$
is equal to (restoring $\hbar$)
\begin{equation}
     \frac{\eta}{s} = \frac{\hbar}{4\pi}\,.
\end{equation}

%%%%%%%%%%%%%%%%%%%%%%%%%%%%%%%%%%%%%%%%%%%%%%%%%%%%%%%%%%%%%%%%%%%%%%%%%%%%
\section{Discussion}
\noindent
Our model (\ref{eq:S1}) can be viewed as an extension of the $d=2$
model of Ref.~\cite{Herzog:2008wg} to arbitrary dimension.
Such a generalization is non-trivial because it involves
both guessing the action and finding the solution to the equations of motion. 
We have not attempted to find a string theory embedding of the action (\ref{eq:S1}),
and we do not know whether black hole solutions (\ref{eq:bh-uv}) uplift to
ten- or eleven-dimensional supergravity for $d>2$.
Nevertheless, we feel that the holographic model (\ref{eq:S1})
is interesting in its own right, in particular because it provides
a window to a novel class of scale invariant $3+1$ dimensional
non-relativistic models with strong quantum fluctuations.
One way to see that the quantum fluctuations in the dual field theory
are strong is to note that the viscosity is small and saturates the
conjectured quantum-mechanical bound of Ref.~\cite{Kovtun:2004de}.
This means that the dual quantum system can not be described as a gas of
weakly interacting quasiparticles because the latter would imply
a parametrically large viscosity to entropy ratio.
Another way to see that interactions are important
is to note that the thermodynamic scaling function $g(x)$
defined by Eq.~(\ref{eq:thermodynamics}) has different form in a
non-interacting gas, and in the present holographic model.
In a non-interacting gas of fermions or bosons in $3+1$ dimensions,
the scaling function is proportional to
$ g(x)\propto \int_0^\infty  {\rm d}u\,u^{3/2}\,(e^{u-x}\pm1)^{-1}$
\cite{LL5},
while in the holographic model $g(x)\propto |x|^{-5/2}$.

The energy per particle
$E/N=\epsilon/n$ in the holographic model is
$$
   \frac{E}{N}=\frac35|\mu|\,,
$$
(in arbitrary dimension, $E/N={|\mu|d}/{(d{+}2)}\,$)
at any temperature, including $T\to0$.
Expressed in terms of the number density, 
the energy per particle is proportional to $n^{-2/7}T^{10/7}$ (in arbitrary
dimension, $\sim n^{-2/(d+4)}T^{2(d+2)/(d+4)}$),
and as a result $E/N\to0$ as $T\to0$ at fixed density.
For unitary fermions at $T=0$, it is conventional to parametrize
the energy per particle as $E/N=\frac35\eF\,\xi$,
where $\eF\sim n^{2/3}$ is the Fermi energy in a
gas of non-interacting fermions.
Lattice simulations \cite{Lee:2005it}
suggest that the thermodynamic scaling function for the
black holes discussed in this paper differs from that of
unitary fermions, in particular giving $\xi$ about $0.4$~\cite{Carlson:2003}.
In the holographic model, on the other hand, the temperature dependence of $E/N$
suggests that $\xi$ is zero in any dimension.
However, it is possible that at lower temperatures in our model
there exist different
backgrounds with the same asymptotics and different thermodynamic properties
which are closer to those of unitary fermions.
This would be natural as the background (\ref{eq:bh-uv}) is invariant
under $v$-translations. Therefore it does not break
particle number symmetry and should not correspond to a superfluid state in
the boundary theory.
It would be interesting to find physical systems with the same
thermodynamic scaling function as in the holographic model.

In this paper, we only considered a holographic model
corresponding to a quantum field theory with the dynamical exponent $z=2$.
It would be interesting to find black holes corresponding to
non-relativistic systems with $z\neq2$ in arbitrary $d$.
Another question to understand is whether
black holes (\ref{eq:bh-uv}) can be embedded in string theory. 
Finally, we see that for the non-relativistic field theories studied here,
the ratio of shear viscosity to entropy density
has a universal value of $\hbar/4\pi$ in any dimension $d\geqslant 2$,
despite the fact that the black hole backgrounds (\ref{eq:bh-uv})
do not satisfy the assumptions
of universality theorems of \cite{Kovtun:2004de,Buchel:2004qq,Benincasa:2006fu}.
This suggests that the universality of shear viscosity can be proven
for a wider class of black hole spacetimes,
and therefore for a wider class of field theories.

\acknowledgments
\noindent
We thank John McGreevy for numerous helpful conversations,
and Chris Herzog, Yusuke Nishida, Krishna Rajagopal and Adam Ritz for comments.
P.K. thanks the MIT Center for Theoretical Physics where part of this work
was completed.
This work was supported in part
by funds provided by the U.S. Department of Energy
(D.O.E.) under cooperative research agreement DE-FG0205ER41360,
by the German Research Foundation (DFG) under grant number Ni 1191/1-1,
and by NSERC of Canada.

%%%%%%%%%%%%%%%%%%%%%%%%%%%%%%%%%%%%%%%%%%%%%%%%%%%%%%%%%%%%%%%

\end{document}